\documentclass[prd,a4paper,twocolumn,showpacs,floatfix]{revtex4-1}

    \usepackage[utf8]{inputenc}
    \usepackage[T1]{fontenc}

    \usepackage{amssymb}
    \usepackage{amsmath}
    \usepackage{amsbsy}
    \usepackage{bm}
    \usepackage{graphicx}
 \usepackage{xcolor}

\DeclareMathOperator\erf{erf}

\begin{document}
\title{Parametric Resonance in Quantum Electrodynamics Vacuum Birefringence}

\author{Ariel Arza$^{1}$, Ricardo Gabriel Elias$^{1, 2}$}

\affiliation{$^{1}$ Departamento de F\'isica, Universidad de Santiago de Chile, Avda. Ecuador 3493, Santiago, Chile}
\affiliation{$^{2}$ CEDENNA, Universidad de Santiago de Chile, Avda. Ecuador 3493, Santiago, Chile.}

\begin{abstract}
Vacuum magnetic birefringence is one of the most interesting non-linear phenomena in quantum electrodynamics because it is a pure photon-photon result of the theory and it directly signalizes the violation of the classical superposition principle of electromagnetic fields in the
full quantum theory. We perform analytical and numerical calculations when an electromagnetic wave interacts with an oscillating external magnetic field. We find that in an ideal cavity, when the external field frequency is around the electromagnetic wave frequency, the normal and parallel components of the wave suffer parametric resonance at different rates, producing a vacuum birefringence effect growing in time. We also study the case where there is no cavity and the oscillating magnetic field is spatially localized in a region of length $L$. In both cases we find also a rotation of the elliptical axis. 
\end{abstract}	


\maketitle

\section{Introduction}

Non-linear effects in the vacuum of quantum electrodynamics (QED) were studied soon after the formulation of QED theory \cite{PhysRev.82.664}, thereby confirming the seminal results obtained even earlier \cite{weisskopf-2024aa,heisenberg-1936aa}. At leading order in a weak field expansion, these non-linear effects arise from microscopic photon-photon scattering processes mediated by an electron-positron loop featuring four couplings to the photon field \cite{PhysRev.83.776,DeTollis1965, ADLER1971599, PhysRevD.3.618, euler-1935aa}, also impacting light propagation in external electromagnetic fields \cite{PhysRev.83.776,DeTollis1965, ADLER1971599, PhysRevD.3.618}, and remaining one of the predictions of QED that has not yet been experimentally corroborated.  The verification of this QED induced effective photon-photon interaction is the major goal of the PVLAS \cite{Bakalov1998, delavalle-2013fg, DellaValle2016}, BMV \cite{refId0, cadene-2014aa} and OVAL \cite{fan-2017aa} experiments. Alternative approaches envision the combination of x-ray free electron and high-power optical lasers \cite{PhysRevD.92.071301, Schlenvoigt2016, PhysRevD.94.013004}. Apart from the QED effect, these experiments can also test other effects and new physics, such as Axion-like particles (ALPs) \cite{maiani-aa, PhysRevLett.59.396, PhysRevD.37.1237}, millicharged Dirac fermions \cite{PhysRevLett.97.140402,PhysRevD.75.035011} and scalar particles \cite{SCHUBERT2000407}.

In the context of QED, it was shown long time ago \cite{PhysRevD.3.618, Baier1967, PhysRevD.2.2341, PhysRevD.2.2341} that the vacuum speed of light in the presence of strong magnetic or electric fields differs from its value $c$ at zero field. Its value -- or equivalently the corresponding refractive index -- depends on the polarization of the light and is different for light polarized parallel and perpendicular to the external field, giving rise to a birefringence phenomenon \cite{PhysRevD.3.618, toll-1952aa, baier-1967aa, PhysRevD.12.1132, Baier1967, PhysRevD.2.2341, IACOPINI1979151} known as vacuum birefringence (VB). Under this context, in order to increase the signals produced by vacuum birefringence effects, large external static magnetic fields and Fabry-Perot cavities have been implemented in optical experiments such as BFRT and PVLAS \cite{PhysRevD.47.3707}.

In this article we study QED vacuum birefringence in an oscillating external magnetic field. We analyze the case of a spatially homogeneous, temporally oscillating magnetic field confined to a conducting cavity, and also the case of a spatially localized, temporally oscillating magnetic field. This article is organized as follows: in the second section we introduce the effective fourth-order Euler–Heisenberg–Weisskopf (EHW) Lagrangian of the theory and the constitutive relations between the fields. In the third section we consider an oscillating magnetic field in a conducting cavity, showing the appearance of the phenomenon of parametric resonance. In the fourth section we study the effects of a temporally oscillating, spatially localized field without cavity. In the last section we summarize and discuss the results.

\section{Euler–Heisenberg–Weisskopf effective Lagrangian and field equations}

As was pointed out by Schwinger \cite{PhysRev.82.664}, non-linear effects  in QED predominate for fields above the critical values for the electric $E=|{\bf E}|$ and magnetic $B=|{\bf B}|$ fields given by $E_{\text{cr}}=m_e^2c^3/q_e\hbar\simeq 1.3\cdot 10^{18}\text{V/m}$ and $B_{\text{cr}}\simeq E_{\text{cr}}/c$, where $m_e$, $q_e$,  $\hbar$ stand for the electron mass, the electron charge and the reduced Planck constant. For fields well under these values, Maxwell equations are enough to describe electromagnetic phenomena. Maxwell equations can be derived from the classical Lagrangian $L_0=\frac{1}{2\mu_0}\left(E^2/c^2-B^2\right)$,  in S.I. units, where $\mu_0$ is the magnetic permeability of the vacuum. For fields below the critical values but big enough to present non-linear effects and varying in distances larger than the Compton wavelength $\lambda_C=2\pi \hbar/m_e c$ (and, equivalently, in times bigger than $\lambda_C/c$) \cite{PhysRevD.91.113002} it should be considered the fourth-order effective Lagrangian derived by Euler, Heisenberg and Weisskopf \cite{euler-1935aa, heisenberg-1936aa, weisskopf-2024aa}
\begin{equation}
L_{\text{EHW}}=L_0+\frac{A_e}{\mu_0}\left[\left(\frac{E^2}{c^2}-B^2\right)^2+\frac{7}{c^2}\left(\bm E\cdot \bm B\right)^2\right],
\label{eq_Lehw}
\end{equation}
where the parameter $A_e$ is defined as:
$$
A_e=\frac{2}{45\mu_0}\frac{\alpha^2\lambdabar^3}{m_ec^2}\simeq 1.32 \cdot 10^{-24} \text{T}^{-2},
$$
where $\lambdabar=\lambda_C /2\pi$ is the reduced Compton wavelength and $\alpha$ is the fine structure constant. The displacement vector $\bm D$ and the magnetizing field $\bm H$ are calculated from the total effective lagrangian Eq. (\ref{eq_Lehw}) using the constitutive relations: 
$$
\bm D=\frac{\partial L}{\partial \bm E}
$$
$$
\bm H=-\frac{\partial L}{\partial \bm B},
$$
giving:
\begin{align}
\bm D=&\epsilon_0\bm E+\epsilon_0 A_e\left[4\left(\frac{E^2}{c^2}-B^2\right)\bm E+14\left(\bm E\cdot\bm B\right)\bm B\right]\nonumber \\ 
\bm H=&\frac{\bm B}{\mu_0}+\frac{A_e}{\mu_0}\left[4\left(\frac{E^2}{c^2}-B^2\right)\bm B-14\left(\frac{\bm E}{c}\cdot\bm B\right)\right],
\label{eqs_DH}
\end{align}
where $\epsilon_0$ is the vacuum permittivity. 

We consider weak perturbations $\bm E, \bm B$ around some known configurations of the fields $\bm E_0, \bm B_0$ that are restricted to the case $ E_0,  B_0\ll  E_{\text{cr}},  B_{\text{cr}}$.
More specifically, to simplify the notation, we use the following
identifications: $\bm E\rightarrow \bm E_0+\bm E$ and $\bm B\rightarrow \bm B_0+\bm B$. In this limit we can linearize the Eqs (\ref{eqs_DH}) around $\bm E_0$ and $\bm B_0$. In order to reproduce the most typical conditions \cite{Bakalov1998} we will consider the case where the external field $\bm E_0$ vanishes, obtaining for Eqs (\ref{eqs_DH}):
\begin{align}
\bm D=&\epsilon_0\bm E+\epsilon_0 A_e\left[-4B_0^2\bm E+14\left(\bm E\cdot\bm B_0\right)\bm B_0\right]\nonumber \\ 
\bm H=&\frac{1}{\mu_0}(\bm B_0 +\bm B)-\frac{4 A_e}{\mu_0}\left[B_0^2\bm B_0+2\left(\bm B_0\cdot \bm B\right)\bm B_0+B_0^2 \bm B\right].
\end{align}
These fields fulfill Maxwell equations in vacuum $\partial_t \bm D=\nabla\times \bm H$ and $\nabla \bm \cdot \bm D=0$ and the equations can be solved when knowing the external fields and the boundary conditions. 

We consider an electromagnetic wave propagating in vacuum along the x-axis and in presence of an external magnetic field in z varying in time as $\bm B_0(t)=B_0(t)\hat z$. In terms of the potentials $\phi$ and $\bm A$, the magnetic and electric fields are defined as $\bm E=-\nabla \phi -\partial_t \bm A$ and $\bm B=\nabla \times \bm A$, respectively. In the transversal or Coulomb gauge ($\nabla\cdot  \bm A=0$) the electric potential is zero in vacuum because the electrical density charge  vanishes \cite{Jackson-1998fk}. We will not consider variations of the fields along the y-z plane and all the dynamics we are interested in is along the longitudinal axis. Under these assumptions the equation $\nabla\cdot \bm D=0$ becomes $B_0\partial_z\partial_t A_z=B_0\partial_t \partial_z A_z=0$ which is identically satisfied because of the arguments previously mentioned. We have also $\nabla^2\bm A=\partial_{xx}\bm A$. The propagation along the x-axis together with the transversal gauge give the condition $A_x=0$ and the Maxwell equation relating $\bm D$ and $\bm H$ becomes two decoupled equations for $A_y, A_z$:
\begin{equation}
\partial_{tt}A_i+\alpha_i(t)\partial_t A_i-\beta_i(t)\partial_{xx}A_i=0, 
\label{Eq_potA}
\end{equation}
for $i=y, z$ and where we have chosen $c=1$, and the definitions $\alpha_y(t)\equiv-\frac{8A_e B_0\dot B_0}{1-4A_e B_0^2}\approx-8A_e B_0\dot B_0$, $\beta_y(t)\equiv\frac{1-12A_e B_0^2}{1-4A_e B_0^2}\approx 1-8A_e B_0^2$, $\alpha_z(t)\equiv\frac{20A_e B_0\dot B_0}{1+10A_e B_0^2}\approx 20A_e B_0\dot B_0$, $\beta_z(t)\equiv\frac{1-4A_e B_0^2}{1+10A_e B_0^2}\approx1-14A_e B_0^2$, where we kept only terms up to
linear order in $A_e B_0^2$. For a time-dependent external magnetic field of the form $B_0(t)=b_0\cos\gamma t$, where $\gamma=2\pi/T$, both equations are of the form
\begin{equation}
\partial_{tt}A_i-l_{i,1}\gamma\sin(2\gamma t)\partial_tA_i-[1-l_{i,2}(1+\cos(2\gamma t))]\partial_{xx}A_i=0, \label{eqA1}
\end{equation}
where $i=y,z$ (no sum over repetead indices). The indices $l_{i,k}$ are given by $l_{y,1}=-l_{y,2}=-4\delta$, $l_{z,1}=10\delta$ and $l_{z,2}=7\delta$, where we have defined the parameter $\delta\equiv A_eb_0^2$, which is small for any realistic magnetic field reachable on earth allowing us to perform all calculations only to first order in $\delta$ throughout this article.

In the time-independent magnetic field case ($\gamma=0$), Eqs. (\ref{eqA1}) show that the speed of light is smaller than $c$ and differs for light polarized parallel and perpendicular to the magnetic field. For the perpendicular and parallel polarized modes of the fields these give us the refractive indices $n_\perp=\sqrt{1/\beta_y}$ and $n_\parallel=\sqrt{1/\beta_z}$, respectivelly:
$$
n_\perp\approx 1+4A_e b_0^2
$$ 
and 
$$
n_\parallel\approx 1+7A_e b_0^2,
$$
this is the known magnetic birefringence $\Delta n=n_\parallel-n_\perp=3A_eb_0^2$ \cite{Baier1967, ADLER1971599, PhysRevD.2.2341}.

In order to explore the effect of an oscillating magnetic field let's consider the case where the field is confined to a conducting cavity and there is an initially linearly polarized electromagnetic wave coming into the cavity. 

\section{Oscillating Magnetic field in a cavity}

\begin{figure}
\centering
\includegraphics[width=.35\textwidth]{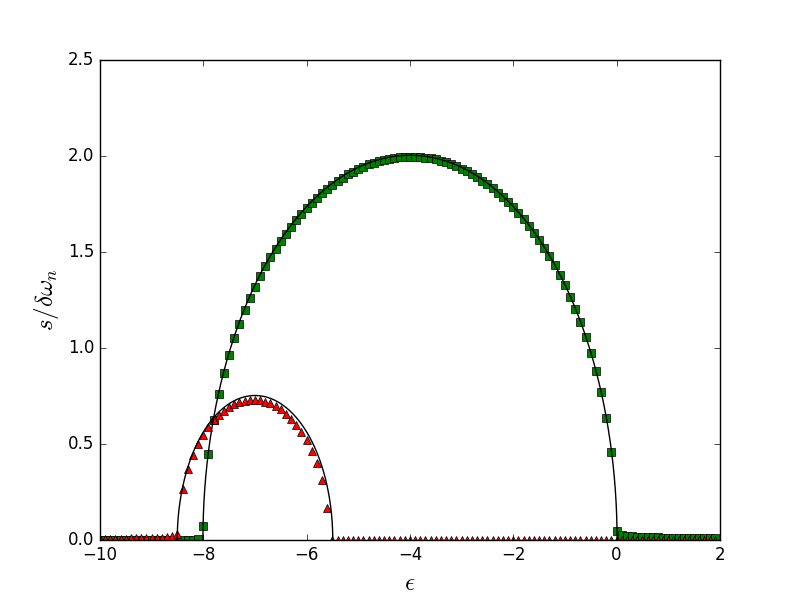}\hfill
\caption{We plot the analytical (black continous line, see formula after Eq. (\ref{eq_epsilon_z})) and numerical exponent (green for $A_y$ and red for $A_z$) $s_i=\lim_{t\to \infty} \frac{\ln E_i(t)}{t}$ versus $\epsilon$ in units of $\delta\omega_n$ (see Eqs. (\ref{eq_epsilon_y}) and (\ref{eq_epsilon_z})).}
\label{fig_evol}
\end{figure}

In this section, we will consider a modulation of period $T$ of the external field. In this way, the coefficients of Eq. (\ref{Eq_potA}) become periodic functions as $\alpha_i(t+T)=\alpha_i(t)$ and $\beta_i(t+T)=\beta_i(t)$, where $\alpha_i(t)\ll 1$ and $\beta_i(t)\approx 1+\beta_{i,s}(t)$, with $\beta_{i,s}(t)\ll 1$.  When considering the field in a conducting cavity of length $L$ and imposing that fields vanish at the boundary of it we are able to decoupling the spatial from the temporal part. The field equations are those of standing waves of the parametrical oscillators type. By doing the factorization $A_i(x,t)=\sum_{n=1}^\infty g_{i,n}(t)\sin\omega_nx$, with $\omega_n=\frac{n\pi}{L}$, $n\in \mathbb{N}$, we impose the condition for standing waves. This give us the resonance frequencies of the cavity. With one of such frequencies the external light wave (usually a laser field) should be tuned \cite{DellaValle2016}. The temporal part for both components of $\bm A$ becomes:
\begin{equation}
\frac{d^2 g_{i,n}}{d t^2}+\alpha_i(t)\frac{dg_{i,n}}{dt}+\omega_{i,n}^2(t)g_{i,n} =0.
\label{eq_g_in}
\end{equation} 
Where we have defined $\alpha_i(t)=-l_{i,1}\gamma\sin2\gamma t$ and $\omega_{i,n}^2(t)=\omega_n^2[1-l_{i,2}(1+\cos2\gamma t)]$. This equation can be transformed using the change of variables $q_i(t)=\exp(D_i(t))g_{i,n}(t)$ with $D_i(t)=\frac{1}{2}\int^t\alpha_i(\tau)d\tau$, giving:
\begin{equation}
\frac{d^2 q_i}{d t^2}+\Omega_{i,n}^2(t)q_i=0,
\label{Eq_q}
\end{equation}
where $\Omega_{i,n}^2(t)=\omega_{i,n}^2(t)-\frac{1}{2}(\frac{\alpha_i^2}{2}+\frac{d\alpha_i}{dt})$. Considering $\delta\ll1$ we have $\Omega_{i,n}^2(t)\approx \omega_{i,n}^2(t)-\frac{1}{2}\frac{d\alpha_i}{dt}=\omega_n^2(1-l_{i,2})-(l_{i,2}\omega_n^2-l_{i,1}\gamma^2)\cos2\gamma t$. The factor 2 in the argument of the cosine aware us that the resonance must be in a neighbourhood of $\omega_n$ and not around twice the natural frequency of the oscillator as in the usual parametric resonance. This fact comes from the squares of magnetic fields in the Eq. (\ref{Eq_potA}). Following Landau \cite{landau-2000aa}, we define $\gamma = \omega_n+ \epsilon/2$. In this way, the relation between detuning $\epsilon$, the period $T$ and the length $L$ of the cavity is given by $\epsilon=2\pi(2/T-n/L)$.  With this definition, we can explicitly write 
\begin{equation}
\frac{d^2 q_i}{d t^2}+\bar\omega_n [1+h_i \cos(2\bar\omega_{i,n} t  +\bar\epsilon_i t)]q_i=0,
\label{eq_landau}
\end{equation} 
where $\bar \omega_{i,n}=\omega_n(1-l_{i,2}/2)$, $\bar \epsilon_i=\epsilon +\omega_n l_{i,2}$ and $h_i=(l_{i,1}-l_{i,2})$. The Eq. (\ref{eq_landau}) is known to have parametric resonance when $-\frac{|h_i|\bar\omega_{i,n}}{2}<\bar\epsilon_i<\frac{|h_i|\bar\omega_{i,n}}{2}$. The $\omega_n$ dependence of $\bar\epsilon_i$ is responsible of the fact that the resonance is not exactly around $\omega_n$ but asymmetrical, which is a particularity of the equations. The resonance region of the detuning $\epsilon$ around the original frequency $\omega_n$ becomes:
\begin{equation}
-\left(l_{i,2}+\frac{|l_{i,1}-l_{i,2}|}{2}\right)\omega_n<\epsilon<-\left(l_{i,2}-\frac{|l_{i,1}-l_{i,2}|}{2}\right)\omega_n.
\end{equation} 
Explicitly, the resonance regions for each component of $\bm A$ are:
\begin{equation}
-8<\frac{\epsilon}{\delta\omega_n}<0, \text{ for $A_y$},
\label{eq_epsilon_y}
\end{equation}
\begin{equation}
-8.5<\frac{\epsilon}{\delta\omega_n}<-5.5, \text{ for $A_z$},
\label{eq_epsilon_z}
\end{equation}
where $\epsilon/\omega_n=2(2L/nT-1)$. For $\epsilon$ inside the resonance region the exponential growth of the fields are giving by $s_i=(1/2)\sqrt{(h_i\bar\omega_{i,n}/2)^2-\bar\epsilon^2}$. The resonance conditions and the exponential growth of the envelope can be tested numerically measuring the behaviour of the fields in time. The predicted regions and exponential growths are in good accord with the numerical simulations of the original field equations in time Eq. (\ref{eq_g_in}), where we can estimate the exponential growth parameter as $s_i=\lim_{t\to \infty} \frac{\ln E_i(t)}{t}$, see Fig. (\ref{fig_evol}). As we can see in the Fig. (\ref{fig_evol}) there is a region where both components resonate although at different rates, being the resonance bigger for $A_y$ which is the component of the potential vector perpendicular to the external field. There is also a tiny region where there is only resonance in $A_z$ and a bigger one where the resonance is only along the $A_y$. In any case, the resonance is more remarkable around $y$.

From now on we will consider the complex notation for fields, in which the actual fields has be understood as the real part of the complex ones \cite{Jackson-1998fk}. In the present case the fields are separable and the temporal part of the electric field is $\bm E(t)=-\frac{\partial\bm A(t)}{\partial t}$. Its components at $t=0$ can be written as $E_i(0)={\cal E}_{0,i}$ 
where ${\cal E}_{0,i}=|{\cal E}_{0,i}|e^{\rm i\varphi_i}$ is the complex amplitude, $\varphi_i$ is the initial phase of each component of the electric field and ${\rm i}$ is the imaginary unit. This phases turn to be the same for both components of the electric field when the initial electromagnetic wave is linearly polarized, which is the case in the present work. So, $\varphi_i=\varphi$. 

The analytical solution for each component of $\bm A$ in the resonance region can be found using a multiscale ansatz supposing that the solution has two different scales of temporal variation: the fast scale is an oscillator of frequency $\gamma$ with an envelope $a(t)$ variating at a bigger temporal scale \cite{landau-2000aa}, in other words $|\dot a(t)|\ll \gamma |a(t)|$. Explicitly, $A_i(t)=a_i(t)e^{{\rm i}\gamma t}$, where $\gamma$ is defined as it was done previous to Eq. (\ref{eq_landau}). Inserting this ansatz in Eq. (\ref{eq_landau}) we find that the solution for the complex amplitude is $a_i(t)=c_{i} \cosh (s_i t)+\frac{1}{2{\rm i} s_i}\left(\epsilon c_{i}-\frac{\omega_n h_i}{2}c_{i}^{*}\right)\sinh(s_i t)$, where $c_{i}$ is a complex constant related with the initial conditions as $c_{i}={\rm i} {\cal E}_{0,i}/\gamma$.  In this way, the explicit form of the envelope is:
\begin{equation}
a_i(t)=\frac{{\cal E}_{0,i}}{\gamma}\left({\rm i}\cosh(s_it)+\frac{1}{2s_i}\left(\epsilon+\frac{\omega_n h_i}{2}\sinh(s_it)\right)  \right).
\end{equation}

This turn to be the main result of this article: the parametric resonance effect with growing at different rate for each component of the field. This effect can be seen when the quantity $s_i t$ is bigger than one.

An important point to taking into account in the case of a non-perfect cavity is the damping in the walls. This damping can be parametrized by a dissipative operator $2\lambda_Q\partial_t$ in the equations of motion, where $\lambda_Q$ is related with the quality factor $Q$ of the cavity by $\lambda_Q=\omega_n/(2Q)$ \cite{landau-2000aa}. This modifies the resonant solution of the electromagnetic field by a factor $e^{-\lambda_Q t}$ \cite{landau-2000aa},  meaning that we have a competition between dissipation and resonance. The resonance in the components $i=y$ or $i=z$ can be seen if
$$
Q>\frac{2}{h_i}.
$$

\subsection{Linear approximation}{}

Let's consider now the limit in which $s_it<1$. We approximate until the first order in $s_it$, explicitly finding the electric field in each direction $i$ as:
\begin{equation}
E_i(x,t)={\cal E}_{0,i}\left[1-\frac{{\rm i}}{2}\left(\epsilon +\frac{h_i\omega_n}{2}e^{-2{\rm i}\varphi}\right) t \right] e^{{\rm i}\omega_n t}\sin(\omega_n x).
\label{eq_E_i}
\end{equation}
From the Stokes parameters $S_0=|E_y|^2+| E_z|^2$,  $S_1=| E_y|^2-| E_z|^2$, $S_2=2\text{Re}( E_y^* E_z)$ and $S_3=2\text{Im}( E_y^* E_z)$, the angle of polarization $\zeta$, the ellipticity angle $\psi$ and the ellipticity $e$ are given by $\tan 2\zeta=S_2/S_1$, $\sin 2\psi= S_3/S_0 $ and $e=\arctan\psi$, respectivelly. For small changes in amplitude and phase, using the relation $\Delta h=h_z-h_y=11 \Delta n/3$, from Eq. (\ref{eq_E_i}) these geometrical factors are:
\begin{equation}
\zeta=\frac{11}{24}\sin2\theta_0\sin 2\varphi \Delta n \omega t
\label{eq_zeta_cav}
\end{equation}
\begin{equation}
e=\frac{11}{24}\sin2\theta_0\cos 2\varphi \Delta n \omega t,
\label{eq_e_cav}
\end{equation}
where $\theta_0$ is the initial angle with the z-axis. As we can see, the rotation of the principal axes of the ellipse is growing in time, which is a particularity of the resonance effect and it is commonly attributed to dichroism. In the context of the biquadratic lagrangian theory we are using here, dichroism was predicted also for electromagnetic waves in presence of strong electric field \cite{PhysRev.136.B1540, 0305-4470-30-18-022}. In this special case dichroism could be attributed to the parametric resonance due to the oscillation of the external field instead of the threshold of pair creation.

Putting the linear approximations Eq. ({\ref{eq_zeta_cav}) and Eq. (\ref{eq_e_cav}}) in function of the  longwave $\lambda=2\pi c/\omega_n$ and $ct\approx N L$, where $N$ is the times of reflections of the wave in the cavity, we found for the maximum value of ellipticity and rotation ($\theta_0=\pi/4, \varphi=0, \pi/4$):
\begin{equation}
e_{\text{max}}=\zeta_{\text{max}}=\frac{11 \pi }{12} \frac{L \Delta n}{\lambda} N,
\label{eq_zeta_e_cav}
\end{equation}
where the ellipticity turns to be very close to the value expected in the PVLAS experiment \cite{delavalle-2013fg, DellaValle2016}. This confirms that in this regime, where the parametric resonance effect is small, we recover the behaviour of the static case. 

\section{Spatially localized oscillating magnetic field}

We now consider the situation in which there is an oscillating magnetic field of frequency $\gamma$, localized in a spatial region ranging from $x=0$ to $x=L$ and no cavity limiting the propagation of the electromagnetic wave. In order to take into account the range of validity of the theory, mentioned in the introduction, following Ref. \cite{karbstein-2015aa}, we consider a spatial magnetic field profile given by a smooth step function where the width of variation is $a\gg \lambda_C$ as:

$$
b_0(x)=\frac{1}{\sqrt{2}} \left[\erf\left(\frac{x}{a}\right)-\erf\left(\frac{x-L}{a}\right)\right]^{\frac{1}{2}}.
$$
This dimensionless magnetic field has the interesting property that $\int_{-\infty}^{\infty} b_0^2(x)dx=L$, that is to say, the energy density remains independent of $a$. In this way, the external magnetic field is $\bm B_0(x,t)= B_0 b_0(x) \cos(\gamma t)\hat z$. The way to proceed is expanding each component of the potential vector as $A_i(x,t)=A_i^{(0)}+A_i^{(1)}+...$  order by order in the parameter $\delta$ and then solving the partial differential equation order by order. Considering the different orders in $\delta$, the linear Eqs. (\ref{Eq_potA}) can be written as
\begin{equation}
{\cal L}(x,t)A_i(x,t)={\cal B}_i(x,t)A_i(x,t),
\label{eq_LAi}
\end{equation}
where ${\cal L}(x,t)=(\partial_{tt}-\partial_{xx})$ is the one-dimensional wave equation operator and the right hand side is first order in $\delta$ with  the operator ${\cal B}_i(x,t)$ defined as: 
\begin{multline}
{\cal B}_i(x,t)=b_0^2(x)[l_{i,1}\gamma \sin(2\gamma t)\partial_t \\ -l_{i,2}(1+\cos(2\gamma t))\partial_{xx}].
\end{multline}

The solution of the zeroth order is a free wave  $A_i^{(0)}(x,t)=a_ie^{{\rm i}\omega( x-t)}$ that we interpret as the incident electromagnetic wave traveling from left to right. At first order the RHS of Eq. (\ref{eq_LAi}) becomes 
\begin{multline}
{\cal B}_i(x,t)A_i^{(0)}(x,t)=b_0^2(x)[\eta_{i,0}e^{{\rm i}(\omega x- \Omega_0 t)}+\eta_{i,+} e^{{\rm i}(\omega x- \Omega_+ t)}\\+ \eta_{i,-}e^{{\rm i}(\omega x- \Omega_- t)}+c.c].
\label{eq_BAi}
\end{multline}
where we define $\eta_{i,0}=a_il_{i,2}\omega^2$, $\eta_{i,{\pm}}=\frac{a_i}{2}\omega(\omega l_{i,2}\pm \gamma l_{i,1})$, $\Omega_0=\omega$, $\Omega_{\pm}=\omega \pm 2\gamma$. 
As was done in previous work studying localized magnetic fields \cite{Adler-2007ab, Gies-2013df, Gies-2015df}, we use the Green function of the one-dimensional wave function operator $G(x-x',t-t')=\frac{1}{2}\theta(t-t'-|x-x'|)$ and the first order of $A_i(x,t)$ becomes:
\begin{equation}
A_i^{(1)}(x,t)=\int dx'\int dt' G(x-x',t-t') {\cal B}_i(x',t')A_i^{(0)}(x',t').
\label{eq_A1_green}
\end{equation} 

The solution is three-fold according to the three regions of propagation. The complete solutions in the external regions where there is no external magnetic field are shown in the Appendix \ref{app_sol_loc}. In the static external magnetic field case the solution for $x<0$ has not relevant contribution compared to the $x>0$ region, which is the usual scenario in magnetic birefringence tests. These results can be found by performing the limit $\gamma\rightarrow 0$ in the Eqs. (\ref{eq_app_Aless}) and (\ref{eq_app_Abigg}). The ellipticity in the static case is 
\begin{equation}
e_{>}=\frac{3}{4}\sin2\theta_0\delta \omega L.
\label{eq_ellip_static}
\end{equation}
On the other hand, in the dynamical case, we found in the region $x<0$, Eq. (\ref{eq_app_Aless}) a backwave (reflected wave) propagating in the oposite sense of the incident light. The first order correction of this wave has resonance in a neighborhood of $\omega$ and its amplitude grows according to the size of $L$. From Eq. (\ref{eq_app_Aless}), for $\gamma=\omega+\epsilon/2$, with $\epsilon\ll \omega$, the relevant term is the one with $\eta_{i,-}$ and it is:
$$
A^{(1)}_{i,<}(x,t)\approx -\frac{\omega^2 h_i a_i e^{-\frac{a^2\epsilon^2}{4}}}{4}\left[\frac{e^{-{\rm i}\epsilon L}-1}{\epsilon(\omega+\epsilon)}\right]e^{{\rm i}(\omega+\epsilon)(x+t)}.
$$
The value of $h_i$ is different for each component of $\bm A$ and it produces the birefringence effect. This leads to an ellipticity given by
\begin{equation}
e_<=\frac{11}{4}\sin2\theta_0\delta\left(\frac{\omega L}{2}\right)^2\frac{\sin\Delta}{\Delta(\Delta+\omega L/2)}\cos\xi(x,t).
\label{eq_e_<}
\end{equation}
Since $\epsilon\ll \omega$ and $\Delta\ll 1$, Eq. (\ref{eq_e_<}) is $e_<\approx \frac{11}{8}\sin2\theta_0\delta \omega L$, which shows that the reflected ellipticity in the time-dependent magnetic field case is as important as the transmitted one in the static case, Eq.(\ref{eq_ellip_static}). This is not found in the constant magnetic field case. On the other hand, we also find a rotation of the ellipse:
\begin{equation}
\zeta_<=\frac{11}{4}\sin2\theta_0\delta\left(\frac{\omega L}{2}\right)^2\frac{\sin\Delta}{\Delta(\Delta+\omega L/2)}\sin\xi(x,t),
\end{equation}
where $\xi(x,t)=(2\omega+\epsilon)x+\epsilon t-\epsilon L/2+2\varphi$ and $\Delta=\epsilon L/2$. The amplitude of these observables have a maximum when $\epsilon=-12/(\omega L^2)$. It is interesting to note that the geometrical factors ellipticity and rotation don't depend on $a$ because its definitions consider the ratio of $S_0, S_1, S_2$ which are all quadratic in fields and share the factors depending on $a$, (see formulae after Eq. (\ref{eq_E_i})).

The transmitted wave has the regular birefringence effect as in the static magnetic field Eq. (\ref{eq_ellip_static}) but reduced by a factor 2. 

\section{Summary and discussion.}

In this paper we have calculated the leading order vacuum birefringence effect predicted by quantum electrodynamics when the external magnetic field is an harmonic time-dependent field. In one spatial dimension and working in the Coulomb gauge, the fields equations for the vector potential become decoupled and can be solved in a spatially localized external magnetic field either confined to a cavity or restricted to an interval of finite length.

In the cavity case, we found that the equations are those of a parametric oscillator, that is to say, the fields perform exponential growth but a different rate in each component. The region of resonance is asymmetrical in the detuning and around $\omega_n$ and not around $2\omega_n$ as in the usual parametric resonance. We compute the ellipticity in the linear case ($s_i t<1$), finding that it is proportional to the reflexions $N$ as in the static case. In the same limit, we also find a rotation of the elliptical axis, which is a new effect. 

When considering the no-cavity and space-localized harmonic magnetic field we found a resonant reflected wave for both components of the vector potential. We have found an ellipticity and rotation of the polarization that is proportional to $L$, where rotation is a particular effect of the dynamical case. For the transmitted wave we found that the ellipticity is reduced by a factor 2 with respect to the static case and no rotation has been found. 

\section{Acknowledgments}

A. A. acknowledges support from USA-1555.

\appendix

\section{Explicit solutions oscillating localized magnetic field.}
\label{app_sol_loc}

At first order $A_i(x,t)\approx A_i^{(0)}+A_i^{(1)}$. The integration in time of the Eq. (\ref{eq_A1_green}) gives a solution with three terms, which have the form
\begin{equation}
{\cal A}^{(1)}_{i,j}(x,t)=\frac{\eta_{i,j}}{2}\int_{-\infty}^{\infty}dx'e^{{\rm i}\omega x'}b_0^2(x)\int_{-\infty}^{t-|x-x'|} dt'e^{-{\rm i}\Omega_j t'},
\end{equation}
where $j$ can be $+,-,0$ as it was defined below Eq. (\ref{eq_BAi}). With these definitions $A_i^{(1)}={\cal A}^{(1)}_{i,0}+{\cal A}^{(1)}_{i,+}+{\cal A}^{(1)}_{i,-}$. Each one of these terms can be simplified as
\begin{equation}
{\cal A}_{i,j}^{(1)}=\frac{{\rm i}\eta_{i,j}}{2\Omega_j}[e^{{\rm i}\Omega_j (x-t)}I_{T,j}(x)+e^{-{\rm i}\Omega_j (x+t)}I_{R,j}(x)],
\label{eq_Aij1}
\end{equation}
where $I_T(x)$ and $I_R(x)$ represent weight functions for the transmitted and the reflected wave and are defined as
$$
I_{T,j}(x)=\int_{-\infty}^{x}dx'e^{{\rm i}(\omega -\Omega_j)x'}b_0^2(x'),
$$
and
$$
I_{R,j}(x)=\int_{x}^{\infty}dx'e^{{\rm i}(\omega+\Omega_j)x'}b_0^2(x').
$$
Indeed, excluding the inner region $0<x<L$, the values of $I_{T,j}(x)$ and $I_{R,j}(x)$ are different from zero only when $x>L$ and $x<0$, respectively. 

\begin{figure}
\centering
\includegraphics[width=.45\textwidth]{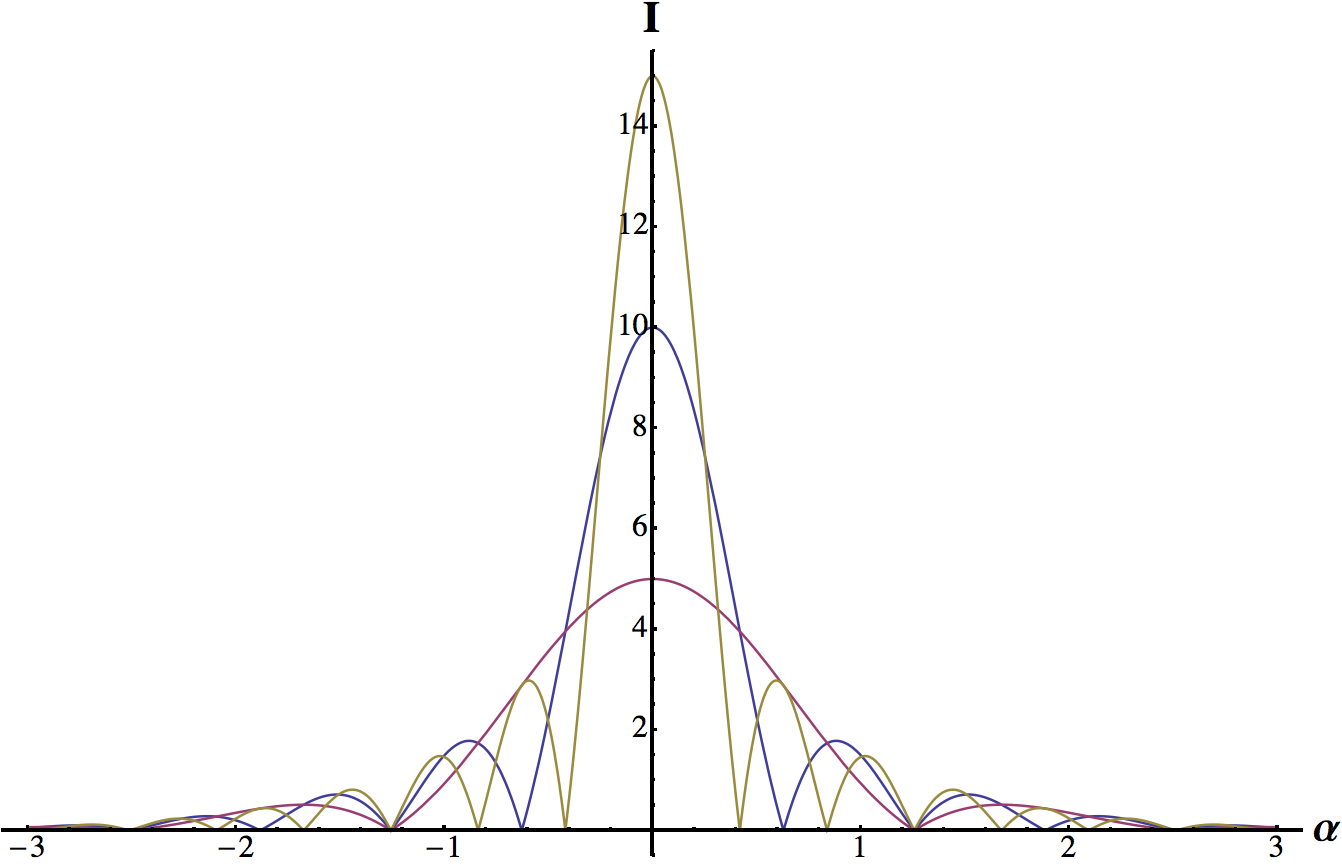}\hfill
\caption{Behaviour of $I_T$ and $I_R$ as function of $\alpha$ in their asymptotic region ($x\rightarrow\infty$ for $I_T$ and $x\rightarrow-\infty$ for $I_R$ for different values of $L=5, 10, 15$. The maximum values of the functions are reached for $\alpha=0$ and are equal to $L$.}
\label{fig_I_alpha}
\end{figure}

In order to capture the important facts of Eq. (\ref{eq_Aij1}) we consider the indefinite integral 
\begin{multline}
I_\alpha(x)=\int dx e^{{\rm i}\alpha x}b_0^2(x)\\=\frac{1}{2\alpha}\left[{\rm i}e^{{\rm i} L\alpha-\frac{a^2\alpha^2}{4}}\erf\left(\frac{2L+{\rm i}a^2\alpha-2x}{2a}\right)\right. \\- {\rm i}e^{{\rm i} \alpha x}\erf\left(\frac{L-x}{a}\right)- {\rm i}e^{{\rm i} \alpha x}\erf\left(\frac{x}{a}\right)\\
 \left. -{\rm i}e^{-\frac{a^2\alpha^2}{4}}\erf\left(\frac{{\rm i} \alpha a}{2}-\frac{x}{a}\right)\right],
\label{eq_I}
\end{multline}
for any $\alpha \in \Re$. This function drops rapidly to a constant value with respect to $x$ outside the magnetic field region, as expected from the localization of the function $b_0(x)$. The numerical evaluation of the functions $I_{T,j}$ and $I_{R,j}$ show that the important contributions come from the region around $\alpha=0$ and their maximum values are proportional to $L$, as is shown in Fig. \ref{fig_I_alpha} for different $L$.

The analytical behaviour of Eq. (\ref{eq_I}) for its asymptotic values is found using the relation $\lim_{x\rightarrow \pm\infty}\erf(x+i c)=\pm 1$, for any $c\in \Re$ which implies 


\begin{equation}
\left. I_{R,j}\right |_{x\rightarrow-\infty}=\left.I_{T,j}\right|_{x\rightarrow\infty}={\rm i}e^{-a^2\alpha^2/4}\left(\frac{1-e^{{\rm i}\alpha L}}{\alpha}\right),
\label{eq_lim_I_1}
\end{equation}
and
\begin{equation}
\left. I_{R,j}\right |_{x\rightarrow\infty}=\left.I_{T,j}\right|_{x\rightarrow-\infty}=0,
\label{eq_lim_I_2}
\end{equation}
which is in accord with the results in Ref. \cite{karbstein-2015aa} The asymptotic value of $I_\alpha(x)$ is reached just at a distance of order $a$ outside the magnetic field localization. For small $a$ we can analytically calculate all the components of the ${\cal A}^{(1)}_{i,j}(x,t)$. Replacing the asymptotic values of $I_\alpha$ in Eq. (\ref{eq_Aij1}) we can explicitly find the reflected and transmitted field. We are mainly interested in the fields out of the magnetic field region.  For $x\leq 0$ the correction becomes:
\begin{multline}
A^{(1)}_{i,<}=\frac{\eta_{i,0}e^{-(a\omega)^2}}{4\omega^2}[e^{2{\rm i}\omega L}-1]e^{-{\rm i}\omega (x+t)}\\
+\frac{\eta_{i,+}e^{-a^2(\omega+\gamma)^2}}{4(\omega+2\gamma)(\omega+\gamma)}[e^{2{\rm i}(\omega+\gamma) L}-1]e^{-{\rm i}(\omega +2\gamma)(x+t)}\\
+\frac{\eta_{i,-}e^{-a^2(\omega-\gamma)^2}}{4(\omega-2\gamma)(\omega-\gamma)}[e^{2{\rm i}(\omega-\gamma) L}-1]e^{-{\rm i}(\omega -2\gamma)(x+t)},
\label{eq_app_Aless}
\end{multline}
while in the $ x > L$ region it yields:
\begin{multline}
A^{(1)}_{i,>}=\eta_{i,0}\frac{ iL}{2\omega}e^{{\rm i}\omega (x-t)}\\
-\frac{\eta_{i,+} e^{-(a\gamma)^2}}{4\gamma(\omega+2\gamma)}[e^{-2{\rm i}\gamma L}-1]e^{{\rm i}(\omega +2\gamma)(x-t)}\\
+\frac{\eta_{i,-}e^{-(a\gamma)^2}}{4\gamma(\omega-2\gamma)}[e^{2{\rm i}\gamma L}-1]e^{{\rm i}(\omega -2\gamma)(x-t)}.\\
\label{eq_app_Abigg}
\end{multline}

These equations can be worked out to find the maximum values of the transmitted and reflected wave (see main text) and they are valid for any external frequencies except in the singular case $\gamma=\omega/2$ where the source has a term with no-temporal dependence. 

In the limit $a\rightarrow0$, the function $b_0(x)\rightarrow\Theta(x)\Theta(L-x)$, in that case Eq. (\ref{eq_lim_I_1}) becomes
\begin{equation}
\left. I'_{T,j}\right |_{x\rightarrow\infty}=\left. I'_{R,j}\right |_{x\rightarrow-\infty}={\rm i}\left(\frac{1-e^{{\rm i}\alpha L}}{\alpha}\right). 
\label{ITR}
\end{equation}
In order to determine the relevance of the parameter $a$, we can define the coefficient  $R=(I'-I)/I'$, which say us how close are the reflected intensities in the discontinuous and continuous magnetic field cases. From Eq. (\ref{ITR}) and the definition of $R$, we find:
\begin{equation}
R=1-e^{-a^2\alpha^2/4}\approx\frac{a^2\alpha^2}{4}. 
\label{R1}
\end{equation}
For instance, for the reflected wave, the important contribution is ${\cal A_-}$, and therefore $\alpha=-\epsilon$. Imposing $R\ll 1$ in Eq. (\ref{R1}) we find that the discontinuous case is a good approximation when the detuning fulfils the relation $\epsilon\ll 1/a$.



\bibliography{VB}

\begin{thebibliography}{40}%
\makeatletter
\providecommand \@ifxundefined [1]{%
 \@ifx{#1\undefined}
}%
\providecommand \@ifnum [1]{%
 \ifnum #1\expandafter \@firstoftwo
 \else \expandafter \@secondoftwo
 \fi
}%
\providecommand \@ifx [1]{%
 \ifx #1\expandafter \@firstoftwo
 \else \expandafter \@secondoftwo
 \fi
}%
\providecommand \natexlab [1]{#1}%
\providecommand \enquote  [1]{``#1''}%
\providecommand \bibnamefont  [1]{#1}%
\providecommand \bibfnamefont [1]{#1}%
\providecommand \citenamefont [1]{#1}%
\providecommand \href@noop [0]{\@secondoftwo}%
\providecommand \href [0]{\begingroup \@sanitize@url \@href}%
\providecommand \@href[1]{\@@startlink{#1}\@@href}%
\providecommand \@@href[1]{\endgroup#1\@@endlink}%
\providecommand \@sanitize@url [0]{\catcode `\\12\catcode `\$12\catcode
  `\&12\catcode `\#12\catcode `\^12\catcode `\_12\catcode `\%12\relax}%
\providecommand \@@startlink[1]{}%
\providecommand \@@endlink[0]{}%
\providecommand \url  [0]{\begingroup\@sanitize@url \@url }%
\providecommand \@url [1]{\endgroup\@href {#1}{\urlprefix }}%
\providecommand \urlprefix  [0]{URL }%
\providecommand \Eprint [0]{\href }%
\providecommand \doibase [0]{http://dx.doi.org/}%
\providecommand \selectlanguage [0]{\@gobble}%
\providecommand \bibinfo  [0]{\@secondoftwo}%
\providecommand \bibfield  [0]{\@secondoftwo}%
\providecommand \translation [1]{[#1]}%
\providecommand \BibitemOpen [0]{}%
\providecommand \bibitemStop [0]{}%
\providecommand \bibitemNoStop [0]{.\EOS\space}%
\providecommand \EOS [0]{\spacefactor3000\relax}%
\providecommand \BibitemShut  [1]{\csname bibitem#1\endcsname}%
\let\auto@bib@innerbib\@empty
\bibitem [{\citenamefont {Schwinger}(1951)}]{PhysRev.82.664}%
  \BibitemOpen
  \bibfield  {author} {\bibinfo {author} {\bibfnamefont {J.}~\bibnamefont
  {Schwinger}},\ }\href {\doibase 10.1103/PhysRev.82.664} {\bibfield  {journal}
  {\bibinfo  {journal} {Phys. Rev.}\ }\textbf {\bibinfo {volume} {82}},\
  \bibinfo {pages} {664} (\bibinfo {year} {1951})}\BibitemShut {NoStop}%
\bibitem [{\citenamefont {Weisskopf}(1936)}]{weisskopf-2024aa}%
  \BibitemOpen
  \bibfield  {author} {\bibinfo {author} {\bibfnamefont {V.~S.}\ \bibnamefont
  {Weisskopf}},\ }\href@noop {} {\bibfield  {journal} {\bibinfo  {journal}
  {Kongelige Danske Videnskabernes Selskab, Mathematisk-fysiske Meddelelser}\
  }\textbf {\bibinfo {volume} {24}} (\bibinfo {year} {1936})}\BibitemShut
  {NoStop}%
\bibitem [{\citenamefont {Heisenberg}\ and\ \citenamefont
  {Euler}(1936)}]{heisenberg-1936aa}%
  \BibitemOpen
  \bibfield  {author} {\bibinfo {author} {\bibfnamefont {W.}~\bibnamefont
  {Heisenberg}}\ and\ \bibinfo {author} {\bibfnamefont {H.}~\bibnamefont
  {Euler}},\ }\href@noop {} {\bibfield  {journal} {\bibinfo  {journal} {Z.
  Phys.}\ }\textbf {\bibinfo {volume} {98}} (\bibinfo {year}
  {1936})}\BibitemShut {NoStop}%
\bibitem [{\citenamefont {Karplus}\ and\ \citenamefont
  {Neuman}(1951)}]{PhysRev.83.776}%
  \BibitemOpen
  \bibfield  {author} {\bibinfo {author} {\bibfnamefont {R.}~\bibnamefont
  {Karplus}}\ and\ \bibinfo {author} {\bibfnamefont {M.}~\bibnamefont
  {Neuman}},\ }\href {\doibase 10.1103/PhysRev.83.776} {\bibfield  {journal}
  {\bibinfo  {journal} {Phys. Rev.}\ }\textbf {\bibinfo {volume} {83}},\
  \bibinfo {pages} {776} (\bibinfo {year} {1951})}\BibitemShut {NoStop}%
\bibitem [{\citenamefont {De~Tollis}(1965)}]{DeTollis1965}%
  \BibitemOpen
  \bibfield  {author} {\bibinfo {author} {\bibfnamefont {B.}~\bibnamefont
  {De~Tollis}},\ }\href {\doibase 10.1007/BF02735534} {\bibfield  {journal}
  {\bibinfo  {journal} {Il Nuovo Cimento (1955-1965)}\ }\textbf {\bibinfo
  {volume} {35}},\ \bibinfo {pages} {1182} (\bibinfo {year}
  {1965})}\BibitemShut {NoStop}%
\bibitem [{\citenamefont {Adler}(1971)}]{ADLER1971599}%
  \BibitemOpen
  \bibfield  {author} {\bibinfo {author} {\bibfnamefont {S.~L.}\ \bibnamefont
  {Adler}},\ }\href {\doibase http://dx.doi.org/10.1016/0003-4916(71)90154-0}
  {\bibfield  {journal} {\bibinfo  {journal} {Annals of Physics}\ }\textbf
  {\bibinfo {volume} {67}},\ \bibinfo {pages} {599 } (\bibinfo {year}
  {1971})}\BibitemShut {NoStop}%
\bibitem [{\citenamefont {Brezin}\ and\ \citenamefont
  {Itzykson}(1971)}]{PhysRevD.3.618}%
  \BibitemOpen
  \bibfield  {author} {\bibinfo {author} {\bibfnamefont {E.}~\bibnamefont
  {Brezin}}\ and\ \bibinfo {author} {\bibfnamefont {C.}~\bibnamefont
  {Itzykson}},\ }\href {\doibase 10.1103/PhysRevD.3.618} {\bibfield  {journal}
  {\bibinfo  {journal} {Phys. Rev. D}\ }\textbf {\bibinfo {volume} {3}},\
  \bibinfo {pages} {618} (\bibinfo {year} {1971})}\BibitemShut {NoStop}%
\bibitem [{\citenamefont {Euler}\ and\ \citenamefont
  {K{\"o}ckel}(1935)}]{euler-1935aa}%
  \BibitemOpen
  \bibfield  {author} {\bibinfo {author} {\bibfnamefont {H.}~\bibnamefont
  {Euler}}\ and\ \bibinfo {author} {\bibfnamefont {B.}~\bibnamefont
  {K{\"o}ckel}},\ }\href@noop {} {\bibfield  {journal} {\bibinfo  {journal}
  {Naturwiss}\ } (\bibinfo {year} {1935})}\BibitemShut {NoStop}%
\bibitem [{\citenamefont {Bakalov}\ \emph {et~al.}(1998)\citenamefont
  {Bakalov}, \citenamefont {Brandi}, \citenamefont {Cantatore}, \citenamefont
  {Carugno}, \citenamefont {Carusotto}, \citenamefont {Della~Valle},
  \citenamefont {De~Riva}, \citenamefont {Gastaldi}, \citenamefont {Iacopini},
  \citenamefont {Micossi}, \citenamefont {Milotti}, \citenamefont {Onofrio},
  \citenamefont {Pengo}, \citenamefont {Perrone}, \citenamefont {Petrucci},
  \citenamefont {Polacco}, \citenamefont {Rizzo}, \citenamefont {Ruoso},
  \citenamefont {Zavattini},\ and\ \citenamefont {Zavattini}}]{Bakalov1998}%
  \BibitemOpen
  \bibfield  {author} {\bibinfo {author} {\bibfnamefont {D.}~\bibnamefont
  {Bakalov}}, \bibinfo {author} {\bibfnamefont {F.}~\bibnamefont {Brandi}},
  \bibinfo {author} {\bibfnamefont {G.}~\bibnamefont {Cantatore}}, \bibinfo
  {author} {\bibfnamefont {G.}~\bibnamefont {Carugno}}, \bibinfo {author}
  {\bibfnamefont {S.}~\bibnamefont {Carusotto}}, \bibinfo {author}
  {\bibfnamefont {F.}~\bibnamefont {Della~Valle}}, \bibinfo {author}
  {\bibfnamefont {A.}~\bibnamefont {De~Riva}}, \bibinfo {author} {\bibfnamefont
  {U.}~\bibnamefont {Gastaldi}}, \bibinfo {author} {\bibfnamefont
  {E.}~\bibnamefont {Iacopini}}, \bibinfo {author} {\bibfnamefont
  {P.}~\bibnamefont {Micossi}}, \bibinfo {author} {\bibfnamefont
  {E.}~\bibnamefont {Milotti}}, \bibinfo {author} {\bibfnamefont
  {R.}~\bibnamefont {Onofrio}}, \bibinfo {author} {\bibfnamefont
  {R.}~\bibnamefont {Pengo}}, \bibinfo {author} {\bibfnamefont
  {F.}~\bibnamefont {Perrone}}, \bibinfo {author} {\bibfnamefont
  {G.}~\bibnamefont {Petrucci}}, \bibinfo {author} {\bibfnamefont
  {E.}~\bibnamefont {Polacco}}, \bibinfo {author} {\bibfnamefont
  {C.}~\bibnamefont {Rizzo}}, \bibinfo {author} {\bibfnamefont
  {G.}~\bibnamefont {Ruoso}}, \bibinfo {author} {\bibfnamefont
  {E.}~\bibnamefont {Zavattini}}, \ and\ \bibinfo {author} {\bibfnamefont
  {G.}~\bibnamefont {Zavattini}},\ }\href {\doibase 10.1023/A:1012610102642}
  {\bibfield  {journal} {\bibinfo  {journal} {Hyperfine Interactions}\ }\textbf
  {\bibinfo {volume} {114}},\ \bibinfo {pages} {103} (\bibinfo {year}
  {1998})}\BibitemShut {NoStop}%
\bibitem [{\citenamefont {Valle}\ \emph {et~al.}(2013)\citenamefont {Valle},
  \citenamefont {Gastaldi}, \citenamefont {Messineo}, \citenamefont {Milotti},
  \citenamefont {Pengo}, \citenamefont {Piemontese}, \citenamefont {Ruoso},\
  and\ \citenamefont {Zavattini}}]{delavalle-2013fg}%
  \BibitemOpen
  \bibfield  {author} {\bibinfo {author} {\bibfnamefont {F.~D.}\ \bibnamefont
  {Valle}}, \bibinfo {author} {\bibfnamefont {U.}~\bibnamefont {Gastaldi}},
  \bibinfo {author} {\bibfnamefont {G.}~\bibnamefont {Messineo}}, \bibinfo
  {author} {\bibfnamefont {E.}~\bibnamefont {Milotti}}, \bibinfo {author}
  {\bibfnamefont {R.}~\bibnamefont {Pengo}}, \bibinfo {author} {\bibfnamefont
  {L.}~\bibnamefont {Piemontese}}, \bibinfo {author} {\bibfnamefont
  {G.}~\bibnamefont {Ruoso}}, \ and\ \bibinfo {author} {\bibfnamefont
  {G.}~\bibnamefont {Zavattini}},\ }\href
  {http://stacks.iop.org/1367-2630/15/i=5/a=053026} {\bibfield  {journal}
  {\bibinfo  {journal} {New Journal of Physics}\ }\textbf {\bibinfo {volume}
  {15}},\ \bibinfo {pages} {053026} (\bibinfo {year} {2013})}\BibitemShut
  {NoStop}%
\bibitem [{\citenamefont {Della~Valle}\ \emph {et~al.}(2016)\citenamefont
  {Della~Valle}, \citenamefont {Ejlli}, \citenamefont {Gastaldi}, \citenamefont
  {Messineo}, \citenamefont {Milotti}, \citenamefont {Pengo}, \citenamefont
  {Ruoso},\ and\ \citenamefont {Zavattini}}]{DellaValle2016}%
  \BibitemOpen
  \bibfield  {author} {\bibinfo {author} {\bibfnamefont {F.}~\bibnamefont
  {Della~Valle}}, \bibinfo {author} {\bibfnamefont {A.}~\bibnamefont {Ejlli}},
  \bibinfo {author} {\bibfnamefont {U.}~\bibnamefont {Gastaldi}}, \bibinfo
  {author} {\bibfnamefont {G.}~\bibnamefont {Messineo}}, \bibinfo {author}
  {\bibfnamefont {E.}~\bibnamefont {Milotti}}, \bibinfo {author} {\bibfnamefont
  {R.}~\bibnamefont {Pengo}}, \bibinfo {author} {\bibfnamefont
  {G.}~\bibnamefont {Ruoso}}, \ and\ \bibinfo {author} {\bibfnamefont
  {G.}~\bibnamefont {Zavattini}},\ }\href {\doibase
  10.1140/epjc/s10052-015-3869-8} {\bibfield  {journal} {\bibinfo  {journal}
  {The European Physical Journal C}\ }\textbf {\bibinfo {volume} {76}},\
  \bibinfo {pages} {24} (\bibinfo {year} {2016})}\BibitemShut {NoStop}%
\bibitem [{\citenamefont {{Battesti, R.}}\ \emph {et~al.}(2008)\citenamefont
  {{Battesti, R.}}, \citenamefont {{Pinto Da Souza, B.}}, \citenamefont
  {{Batut, S.}}, \citenamefont {{Robilliard, C.}}, \citenamefont {{Bailly,
  G.}}, \citenamefont {{Michel, C.}}, \citenamefont {{Nardone, M.}},
  \citenamefont {{Pinard, L.}}, \citenamefont {{Portugall, O.}}, \citenamefont
  {{Tr{\'e}nec, G.}}, \citenamefont {{Mackowski, J.-M.}}, \citenamefont
  {{Rikken, G. L.J.A.}}, \citenamefont {{Vigu{\'e}, J.}},\ and\ \citenamefont
  {{Rizzo, C.}}}]{refId0}%
  \BibitemOpen
  \bibfield  {author} {\bibinfo {author} {\bibnamefont {{Battesti, R.}}},
  \bibinfo {author} {\bibnamefont {{Pinto Da Souza, B.}}}, \bibinfo {author}
  {\bibnamefont {{Batut, S.}}}, \bibinfo {author} {\bibnamefont {{Robilliard,
  C.}}}, \bibinfo {author} {\bibnamefont {{Bailly, G.}}}, \bibinfo {author}
  {\bibnamefont {{Michel, C.}}}, \bibinfo {author} {\bibnamefont {{Nardone,
  M.}}}, \bibinfo {author} {\bibnamefont {{Pinard, L.}}}, \bibinfo {author}
  {\bibnamefont {{Portugall, O.}}}, \bibinfo {author} {\bibnamefont
  {{Tr{\'e}nec, G.}}}, \bibinfo {author} {\bibnamefont {{Mackowski, J.-M.}}},
  \bibinfo {author} {\bibnamefont {{Rikken, G. L.J.A.}}}, \bibinfo {author}
  {\bibnamefont {{Vigu{\'e}, J.}}}, \ and\ \bibinfo {author} {\bibnamefont
  {{Rizzo, C.}}},\ }\href {\doibase 10.1140/epjd/e2007-00306-3} {\bibfield
  {journal} {\bibinfo  {journal} {Eur. Phys. J. D}\ }\textbf {\bibinfo {volume}
  {46}},\ \bibinfo {pages} {323} (\bibinfo {year} {2008})}\BibitemShut
  {NoStop}%
\bibitem [{\citenamefont {Cad{\`e}ne}\ \emph {et~al.}(2014)\citenamefont
  {Cad{\`e}ne}, \citenamefont {Berceau}, \citenamefont {Fouch{\'e}},
  \citenamefont {Battesti},\ and\ \citenamefont {Rizzo}}]{cadene-2014aa}%
  \BibitemOpen
  \bibfield  {author} {\bibinfo {author} {\bibfnamefont {A.}~\bibnamefont
  {Cad{\`e}ne}}, \bibinfo {author} {\bibfnamefont {P.}~\bibnamefont {Berceau}},
  \bibinfo {author} {\bibfnamefont {M.}~\bibnamefont {Fouch{\'e}}}, \bibinfo
  {author} {\bibfnamefont {R.}~\bibnamefont {Battesti}}, \ and\ \bibinfo
  {author} {\bibfnamefont {C.}~\bibnamefont {Rizzo}},\ }\href {\doibase
  10.1140/epjd/e2013-40725-9} {\bibfield  {journal} {\bibinfo  {journal} {The
  European Physical Journal D}\ }\textbf {\bibinfo {volume} {68}},\ \bibinfo
  {pages} {16} (\bibinfo {year} {2014})}\BibitemShut {NoStop}%
\bibitem [{\citenamefont {Fan}\ \emph {et~al.}(2017)\citenamefont {Fan},
  \citenamefont {Kamioka}, \citenamefont {Inada}, \citenamefont {Yamazaki},
  \citenamefont {Namba}, \citenamefont {Asai}, \citenamefont {Omachi},
  \citenamefont {Yoshioka}, \citenamefont {Kuwata-Gonokami}, \citenamefont
  {Matsuo}, \citenamefont {Kawaguchi}, \citenamefont {Kindo},\ and\
  \citenamefont {Nojiri}}]{fan-2017aa}%
  \BibitemOpen
  \bibfield  {author} {\bibinfo {author} {\bibfnamefont {X.}~\bibnamefont
  {Fan}}, \bibinfo {author} {\bibfnamefont {S.}~\bibnamefont {Kamioka}},
  \bibinfo {author} {\bibfnamefont {T.}~\bibnamefont {Inada}}, \bibinfo
  {author} {\bibfnamefont {T.}~\bibnamefont {Yamazaki}}, \bibinfo {author}
  {\bibfnamefont {T.}~\bibnamefont {Namba}}, \bibinfo {author} {\bibfnamefont
  {S.}~\bibnamefont {Asai}}, \bibinfo {author} {\bibfnamefont {J.}~\bibnamefont
  {Omachi}}, \bibinfo {author} {\bibfnamefont {K.}~\bibnamefont {Yoshioka}},
  \bibinfo {author} {\bibfnamefont {M.}~\bibnamefont {Kuwata-Gonokami}},
  \bibinfo {author} {\bibfnamefont {A.}~\bibnamefont {Matsuo}}, \bibinfo
  {author} {\bibfnamefont {K.}~\bibnamefont {Kawaguchi}}, \bibinfo {author}
  {\bibfnamefont {K.}~\bibnamefont {Kindo}}, \ and\ \bibinfo {author}
  {\bibfnamefont {H.}~\bibnamefont {Nojiri}},\ }\href {\doibase
  10.1140/epjd/e2017-80290-7} {\bibfield  {journal} {\bibinfo  {journal} {The
  European Physical Journal D}\ }\textbf {\bibinfo {volume} {71}},\ \bibinfo
  {pages} {308} (\bibinfo {year} {2017})}\BibitemShut {NoStop}%
\bibitem [{\citenamefont {Karbstein}\ \emph {et~al.}(2015)\citenamefont
  {Karbstein}, \citenamefont {Gies}, \citenamefont {Reuter},\ and\
  \citenamefont {Zepf}}]{PhysRevD.92.071301}%
  \BibitemOpen
  \bibfield  {author} {\bibinfo {author} {\bibfnamefont {F.}~\bibnamefont
  {Karbstein}}, \bibinfo {author} {\bibfnamefont {H.}~\bibnamefont {Gies}},
  \bibinfo {author} {\bibfnamefont {M.}~\bibnamefont {Reuter}}, \ and\ \bibinfo
  {author} {\bibfnamefont {M.}~\bibnamefont {Zepf}},\ }\href {\doibase
  10.1103/PhysRevD.92.071301} {\bibfield  {journal} {\bibinfo  {journal} {Phys.
  Rev. D}\ }\textbf {\bibinfo {volume} {92}},\ \bibinfo {pages} {071301}
  (\bibinfo {year} {2015})}\BibitemShut {NoStop}%
\bibitem [{\citenamefont {Schlenvoigt}\ \emph {et~al.}(2016)\citenamefont
  {Schlenvoigt}, \citenamefont {Heinzl}, \citenamefont {Schramm}, \citenamefont
  {Cowan},\ and\ \citenamefont {Sauerbrey}}]{Schlenvoigt2016}%
  \BibitemOpen
  \bibfield  {author} {\bibinfo {author} {\bibfnamefont {H.-P.}\ \bibnamefont
  {Schlenvoigt}}, \bibinfo {author} {\bibfnamefont {T.}~\bibnamefont {Heinzl}},
  \bibinfo {author} {\bibfnamefont {U.}~\bibnamefont {Schramm}}, \bibinfo
  {author} {\bibfnamefont {T.~E.}\ \bibnamefont {Cowan}}, \ and\ \bibinfo
  {author} {\bibfnamefont {R.}~\bibnamefont {Sauerbrey}},\ }\href
  {http://stacks.iop.org/1402-4896/91/i=2/a=023010} {\bibfield  {journal}
  {\bibinfo  {journal} {Physica Scripta}\ }\textbf {\bibinfo {volume} {91}},\
  \bibinfo {pages} {023010} (\bibinfo {year} {2016})}\BibitemShut {NoStop}%
\bibitem [{\citenamefont {Karbstein}\ and\ \citenamefont
  {Sundqvist}(2016)}]{PhysRevD.94.013004}%
  \BibitemOpen
  \bibfield  {author} {\bibinfo {author} {\bibfnamefont {F.}~\bibnamefont
  {Karbstein}}\ and\ \bibinfo {author} {\bibfnamefont {C.}~\bibnamefont
  {Sundqvist}},\ }\href {\doibase 10.1103/PhysRevD.94.013004} {\bibfield
  {journal} {\bibinfo  {journal} {Phys. Rev. D}\ }\textbf {\bibinfo {volume}
  {94}},\ \bibinfo {pages} {013004} (\bibinfo {year} {2016})}\BibitemShut
  {NoStop}%
\bibitem [{\citenamefont {Maiani}\ \emph {et~al.}(1986)\citenamefont {Maiani},
  \citenamefont {Petronzio},\ and\ \citenamefont {Zavattini}}]{maiani-aa}%
  \BibitemOpen
  \bibfield  {author} {\bibinfo {author} {\bibfnamefont {L.}~\bibnamefont
  {Maiani}}, \bibinfo {author} {\bibfnamefont {R.}~\bibnamefont {Petronzio}}, \
  and\ \bibinfo {author} {\bibfnamefont {E.}~\bibnamefont {Zavattini}},\ }\href
  {\doibase http://dx.doi.org/10.1016/0370-2693(86)90869-5} {\bibfield
  {journal} {\bibinfo  {journal} {Physics Letters B}\ }\textbf {\bibinfo
  {volume} {175}},\ \bibinfo {pages} {359 } (\bibinfo {year}
  {1986})}\BibitemShut {NoStop}%
\bibitem [{\citenamefont {Gasperini}(1987)}]{PhysRevLett.59.396}%
  \BibitemOpen
  \bibfield  {author} {\bibinfo {author} {\bibfnamefont {M.}~\bibnamefont
  {Gasperini}},\ }\href {\doibase 10.1103/PhysRevLett.59.396} {\bibfield
  {journal} {\bibinfo  {journal} {Phys. Rev. Lett.}\ }\textbf {\bibinfo
  {volume} {59}},\ \bibinfo {pages} {396} (\bibinfo {year} {1987})}\BibitemShut
  {NoStop}%
\bibitem [{\citenamefont {Raffelt}\ and\ \citenamefont
  {Stodolsky}(1988)}]{PhysRevD.37.1237}%
  \BibitemOpen
  \bibfield  {author} {\bibinfo {author} {\bibfnamefont {G.}~\bibnamefont
  {Raffelt}}\ and\ \bibinfo {author} {\bibfnamefont {L.}~\bibnamefont
  {Stodolsky}},\ }\href {\doibase 10.1103/PhysRevD.37.1237} {\bibfield
  {journal} {\bibinfo  {journal} {Phys. Rev. D}\ }\textbf {\bibinfo {volume}
  {37}},\ \bibinfo {pages} {1237} (\bibinfo {year} {1988})}\BibitemShut
  {NoStop}%
\bibitem [{\citenamefont {Gies}\ \emph {et~al.}(2006)\citenamefont {Gies},
  \citenamefont {Jaeckel},\ and\ \citenamefont
  {Ringwald}}]{PhysRevLett.97.140402}%
  \BibitemOpen
  \bibfield  {author} {\bibinfo {author} {\bibfnamefont {H.}~\bibnamefont
  {Gies}}, \bibinfo {author} {\bibfnamefont {J.}~\bibnamefont {Jaeckel}}, \
  and\ \bibinfo {author} {\bibfnamefont {A.}~\bibnamefont {Ringwald}},\ }\href
  {\doibase 10.1103/PhysRevLett.97.140402} {\bibfield  {journal} {\bibinfo
  {journal} {Phys. Rev. Lett.}\ }\textbf {\bibinfo {volume} {97}},\ \bibinfo
  {pages} {140402} (\bibinfo {year} {2006})}\BibitemShut {NoStop}%
\bibitem [{\citenamefont {Ahlers}\ \emph {et~al.}(2007)\citenamefont {Ahlers},
  \citenamefont {Gies}, \citenamefont {Jaeckel},\ and\ \citenamefont
  {Ringwald}}]{PhysRevD.75.035011}%
  \BibitemOpen
  \bibfield  {author} {\bibinfo {author} {\bibfnamefont {M.}~\bibnamefont
  {Ahlers}}, \bibinfo {author} {\bibfnamefont {H.}~\bibnamefont {Gies}},
  \bibinfo {author} {\bibfnamefont {J.}~\bibnamefont {Jaeckel}}, \ and\
  \bibinfo {author} {\bibfnamefont {A.}~\bibnamefont {Ringwald}},\ }\href
  {\doibase 10.1103/PhysRevD.75.035011} {\bibfield  {journal} {\bibinfo
  {journal} {Phys. Rev. D}\ }\textbf {\bibinfo {volume} {75}},\ \bibinfo
  {pages} {035011} (\bibinfo {year} {2007})}\BibitemShut {NoStop}%
\bibitem [{\citenamefont {Schubert}(2000)}]{SCHUBERT2000407}%
  \BibitemOpen
  \bibfield  {author} {\bibinfo {author} {\bibfnamefont {C.}~\bibnamefont
  {Schubert}},\ }\href {\doibase
  http://dx.doi.org/10.1016/S0550-3213(00)00423-5} {\bibfield  {journal}
  {\bibinfo  {journal} {Nuclear Physics B}\ }\textbf {\bibinfo {volume}
  {585}},\ \bibinfo {pages} {407 } (\bibinfo {year} {2000})}\BibitemShut
  {NoStop}%
\bibitem [{\citenamefont {Baier}\ and\ \citenamefont
  {Breitenlohner}(1967{\natexlab{a}})}]{Baier1967}%
  \BibitemOpen
  \bibfield  {author} {\bibinfo {author} {\bibfnamefont {R.}~\bibnamefont
  {Baier}}\ and\ \bibinfo {author} {\bibfnamefont {P.}~\bibnamefont
  {Breitenlohner}},\ }\href {\doibase 10.1007/BF02712312} {\bibfield  {journal}
  {\bibinfo  {journal} {Il Nuovo Cimento B (1965-1970)}\ }\textbf {\bibinfo
  {volume} {47}},\ \bibinfo {pages} {117} (\bibinfo {year}
  {1967}{\natexlab{a}})}\BibitemShut {NoStop}%
\bibitem [{\citenamefont {Bialynicka-Birula}\ and\ \citenamefont
  {Bialynicki-Birula}(1970)}]{PhysRevD.2.2341}%
  \BibitemOpen
  \bibfield  {author} {\bibinfo {author} {\bibfnamefont {Z.}~\bibnamefont
  {Bialynicka-Birula}}\ and\ \bibinfo {author} {\bibfnamefont {I.}~\bibnamefont
  {Bialynicki-Birula}},\ }\href {\doibase 10.1103/PhysRevD.2.2341} {\bibfield
  {journal} {\bibinfo  {journal} {Phys. Rev. D}\ }\textbf {\bibinfo {volume}
  {2}},\ \bibinfo {pages} {2341} (\bibinfo {year} {1970})}\BibitemShut
  {NoStop}%
\bibitem [{\citenamefont {Toll}(1952)}]{toll-1952aa}%
  \BibitemOpen
  \bibfield  {author} {\bibinfo {author} {\bibfnamefont {J.~S.}\ \bibnamefont
  {Toll}},\ }\href@noop {} {\enquote {\bibinfo {title} {The dispersion relation
  for light and its application to problems involving electron pairs},}\ }
  (\bibinfo {year} {1952}),\ \bibinfo {note} {ph.D. dissertation, Princeton
  University}\BibitemShut {NoStop}%
\bibitem [{\citenamefont {Baier}\ and\ \citenamefont
  {Breitenlohner}(1967{\natexlab{b}})}]{baier-1967aa}%
  \BibitemOpen
  \bibfield  {author} {\bibinfo {author} {\bibfnamefont {R.}~\bibnamefont
  {Baier}}\ and\ \bibinfo {author} {\bibfnamefont {P.}~\bibnamefont
  {Breitenlohner}},\ }\href@noop {} {\bibfield  {journal} {\bibinfo  {journal}
  {Acta Phys. Austriaca}\ }\textbf {\bibinfo {volume} {25}},\ \bibinfo {pages}
  {212} (\bibinfo {year} {1967}{\natexlab{b}})}\BibitemShut {NoStop}%
\bibitem [{\citenamefont {Tsai}\ and\ \citenamefont
  {Erber}(1975)}]{PhysRevD.12.1132}%
  \BibitemOpen
  \bibfield  {author} {\bibinfo {author} {\bibfnamefont {W.-y.}\ \bibnamefont
  {Tsai}}\ and\ \bibinfo {author} {\bibfnamefont {T.}~\bibnamefont {Erber}},\
  }\href {\doibase 10.1103/PhysRevD.12.1132} {\bibfield  {journal} {\bibinfo
  {journal} {Phys. Rev. D}\ }\textbf {\bibinfo {volume} {12}},\ \bibinfo
  {pages} {1132} (\bibinfo {year} {1975})}\BibitemShut {NoStop}%
\bibitem [{\citenamefont {Iacopini}\ and\ \citenamefont
  {Zavattini}(1979)}]{IACOPINI1979151}%
  \BibitemOpen
  \bibfield  {author} {\bibinfo {author} {\bibfnamefont {E.}~\bibnamefont
  {Iacopini}}\ and\ \bibinfo {author} {\bibfnamefont {E.}~\bibnamefont
  {Zavattini}},\ }\href {\doibase
  http://dx.doi.org/10.1016/0370-2693(79)90797-4} {\bibfield  {journal}
  {\bibinfo  {journal} {Physics Letters B}\ }\textbf {\bibinfo {volume} {85}},\
  \bibinfo {pages} {151 } (\bibinfo {year} {1979})}\BibitemShut {NoStop}%
\bibitem [{\citenamefont {Cameron}\ \emph {et~al.}(1993)\citenamefont
  {Cameron}, \citenamefont {Cantatore}, \citenamefont {Melissinos},
  \citenamefont {Ruoso}, \citenamefont {Semertzidis}, \citenamefont {Halama},
  \citenamefont {Lazarus}, \citenamefont {Prodell}, \citenamefont {Nezrick},
  \citenamefont {Rizzo},\ and\ \citenamefont {Zavattini}}]{PhysRevD.47.3707}%
  \BibitemOpen
  \bibfield  {author} {\bibinfo {author} {\bibfnamefont {R.}~\bibnamefont
  {Cameron}}, \bibinfo {author} {\bibfnamefont {G.}~\bibnamefont {Cantatore}},
  \bibinfo {author} {\bibfnamefont {A.~C.}\ \bibnamefont {Melissinos}},
  \bibinfo {author} {\bibfnamefont {G.}~\bibnamefont {Ruoso}}, \bibinfo
  {author} {\bibfnamefont {Y.}~\bibnamefont {Semertzidis}}, \bibinfo {author}
  {\bibfnamefont {H.~J.}\ \bibnamefont {Halama}}, \bibinfo {author}
  {\bibfnamefont {D.~M.}\ \bibnamefont {Lazarus}}, \bibinfo {author}
  {\bibfnamefont {A.~G.}\ \bibnamefont {Prodell}}, \bibinfo {author}
  {\bibfnamefont {F.}~\bibnamefont {Nezrick}}, \bibinfo {author} {\bibfnamefont
  {C.}~\bibnamefont {Rizzo}}, \ and\ \bibinfo {author} {\bibfnamefont
  {E.}~\bibnamefont {Zavattini}},\ }\href {\doibase 10.1103/PhysRevD.47.3707}
  {\bibfield  {journal} {\bibinfo  {journal} {Phys. Rev. D}\ }\textbf {\bibinfo
  {volume} {47}},\ \bibinfo {pages} {3707} (\bibinfo {year}
  {1993})}\BibitemShut {NoStop}%
\bibitem [{\citenamefont {Karbstein}\ and\ \citenamefont
  {Shaisultanov}(2015)}]{PhysRevD.91.113002}%
  \BibitemOpen
  \bibfield  {author} {\bibinfo {author} {\bibfnamefont {F.}~\bibnamefont
  {Karbstein}}\ and\ \bibinfo {author} {\bibfnamefont {R.}~\bibnamefont
  {Shaisultanov}},\ }\href {\doibase 10.1103/PhysRevD.91.113002} {\bibfield
  {journal} {\bibinfo  {journal} {Phys. Rev. D}\ }\textbf {\bibinfo {volume}
  {91}},\ \bibinfo {pages} {113002} (\bibinfo {year} {2015})}\BibitemShut
  {NoStop}%
\bibitem [{\citenamefont {Jackson}(1998)}]{Jackson-1998fk}%
  \BibitemOpen
  \bibfield  {author} {\bibinfo {author} {\bibfnamefont {J.~D.}\ \bibnamefont
  {Jackson}},\ }\href@noop {} {\emph {\bibinfo {title} {Classical
  Electrodynamics}}}\ (\bibinfo  {publisher} {John Wiley and Sons, Inc.},\
  \bibinfo {year} {1998})\BibitemShut {NoStop}%
\bibitem [{\citenamefont {Landau}\ and\ \citenamefont
  {Lifshitz}(2000)}]{landau-2000aa}%
  \BibitemOpen
  \bibfield  {author} {\bibinfo {author} {\bibfnamefont {L.~D.}\ \bibnamefont
  {Landau}}\ and\ \bibinfo {author} {\bibfnamefont {E.~M.}\ \bibnamefont
  {Lifshitz}},\ }\href@noop {} {\emph {\bibinfo {title} {Mechanics}}}\
  (\bibinfo  {publisher} {Butterworth, Heinemann, Oxford},\ \bibinfo {year}
  {2000})\BibitemShut {NoStop}%
\bibitem [{\citenamefont {Jackson}(1962)}]{Jackson-1962aa}%
  \BibitemOpen
  \bibfield  {author} {\bibinfo {author} {\bibfnamefont {J.~D.}\ \bibnamefont
  {Jackson}},\ }\href@noop {} {\emph {\bibinfo {title} {Electrodynamics}}}\
  (\bibinfo  {publisher} {John Wiley and Sons, Inc.},\ \bibinfo {year}
  {1962})\BibitemShut {NoStop}%
\bibitem [{\citenamefont {Klein}\ and\ \citenamefont
  {Nigam}(1964)}]{PhysRev.136.B1540}%
  \BibitemOpen
  \bibfield  {author} {\bibinfo {author} {\bibfnamefont {J.~J.}\ \bibnamefont
  {Klein}}\ and\ \bibinfo {author} {\bibfnamefont {B.~P.}\ \bibnamefont
  {Nigam}},\ }\href {\doibase 10.1103/PhysRev.136.B1540} {\bibfield  {journal}
  {\bibinfo  {journal} {Phys. Rev.}\ }\textbf {\bibinfo {volume} {136}},\
  \bibinfo {pages} {B1540} (\bibinfo {year} {1964})}\BibitemShut {NoStop}%
\bibitem [{\citenamefont {Heyl}\ and\ \citenamefont
  {Hernquist}(1997)}]{0305-4470-30-18-022}%
  \BibitemOpen
  \bibfield  {author} {\bibinfo {author} {\bibfnamefont {J.~S.}\ \bibnamefont
  {Heyl}}\ and\ \bibinfo {author} {\bibfnamefont {L.}~\bibnamefont
  {Hernquist}},\ }\href {http://stacks.iop.org/0305-4470/30/i=18/a=022}
  {\bibfield  {journal} {\bibinfo  {journal} {Journal of Physics A:
  Mathematical and General}\ }\textbf {\bibinfo {volume} {30}},\ \bibinfo
  {pages} {6485} (\bibinfo {year} {1997})}\BibitemShut {NoStop}%
\bibitem [{\citenamefont {Karbstein}(2015)}]{karbstein-2015aa}%
  \BibitemOpen
  \bibfield  {author} {\bibinfo {author} {\bibfnamefont {F.}~\bibnamefont
  {Karbstein}},\ }\href@noop {} {\bibfield  {journal} {\bibinfo  {journal}
  {Arxiv.org}\ ,\ \bibinfo {pages} {1510.03178}} (\bibinfo {year}
  {2015})}\BibitemShut {NoStop}%
\bibitem [{\citenamefont {Adler}(2007)}]{Adler-2007ab}%
  \BibitemOpen
  \bibfield  {author} {\bibinfo {author} {\bibfnamefont {S.~L.}\ \bibnamefont
  {Adler}},\ }\href {http://stacks.iop.org/1751-8121/40/i=5/a=F01} {\bibfield
  {journal} {\bibinfo  {journal} {Journal of Physics A: Mathematical and
  Theoretical}\ }\textbf {\bibinfo {volume} {40}},\ \bibinfo {pages} {F143}
  (\bibinfo {year} {2007})}\BibitemShut {NoStop}%
\bibitem [{\citenamefont {Gies}\ \emph {et~al.}(2013)\citenamefont {Gies},
  \citenamefont {Karbstein},\ and\ \citenamefont {Seegert}}]{Gies-2013df}%
  \BibitemOpen
  \bibfield  {author} {\bibinfo {author} {\bibfnamefont {H.}~\bibnamefont
  {Gies}}, \bibinfo {author} {\bibfnamefont {F.}~\bibnamefont {Karbstein}}, \
  and\ \bibinfo {author} {\bibfnamefont {N.}~\bibnamefont {Seegert}},\ }\href
  {http://stacks.iop.org/1367-2630/15/i=8/a=083002} {\bibfield  {journal}
  {\bibinfo  {journal} {New Journal of Physics}\ }\textbf {\bibinfo {volume}
  {15}},\ \bibinfo {pages} {083002} (\bibinfo {year} {2013})}\BibitemShut
  {NoStop}%
\bibitem [{\citenamefont {Gies}\ \emph {et~al.}(2015)\citenamefont {Gies},
  \citenamefont {Karbstein},\ and\ \citenamefont {Seegert}}]{Gies-2015df}%
  \BibitemOpen
  \bibfield  {author} {\bibinfo {author} {\bibfnamefont {H.}~\bibnamefont
  {Gies}}, \bibinfo {author} {\bibfnamefont {F.}~\bibnamefont {Karbstein}}, \
  and\ \bibinfo {author} {\bibfnamefont {N.}~\bibnamefont {Seegert}},\ }\href
  {http://stacks.iop.org/1367-2630/17/i=4/a=043060} {\bibfield  {journal}
  {\bibinfo  {journal} {New Journal of Physics}\ }\textbf {\bibinfo {volume}
  {17}},\ \bibinfo {pages} {043060} (\bibinfo {year} {2015})}\BibitemShut
  {NoStop}%
\end{thebibliography}%

\end{document}